\input psfig.tex
\documentstyle[12pt,aasms4]{article}

\lefthead{Spaans \& Carollo}
\righthead{Star Formation as a Function of Redshift}

\begin{document}

\title{Star Formation and Metal Production as a Function of Redshift: The
Role of the Multi-Phase ISM}

\author{Marco Spaans and C.~Marcella Carollo\footnote{Hubble Fellow}}
\affil{Department of Physics \& Astronomy, The Johns Hopkins University, Baltimore, MD 21218}

\begin{abstract}

We present models of the cosmological star formation and metal
production history of (proto-)galaxies with varying axis ratios.  More
massive and/or roughly spherical systems reach the
threshold-metallicity for a transition to a multi-phase interstellar
medium earlier than less massive, more flattened systems.  Therefore,
more flattened, lower-mass systems start to form stars actively at
smaller redshifts. A natural explanation is found in the overall
robustness of the interstellar medium against complete expulsion
(blow-away) at high total masses, and in the prevention of metal
enrichment in the outer regions due to axial outflow along the
symmetry axis of a non-spherical proto-galaxy (blow-out).  We suggest
that the observed predominance of spheroidal systems observed at high
redshift, e.g.~in the Hubble Deep Field, is due to this effect: At
$z>$2, roundish (proto-)galaxies with total (dark+baryonic) masses of
$\approx10^{11}$M$_\odot$ and/or the inner spheroidal cores of
similarly massive flattened systems sustain a multi-phase interstellar
medium, and therefore a high star-formation rate, whose magnitude
depends on the fraction of baryonic matter in the systems.
Conversely, the peak at $z\sim 1-2$ in the observed cosmological
metal production rate coincides with the epochs of star formation of
lower mass spheroidals, as well as of massive proto-galactic disks.

\end{abstract}

{\it subject headings}: cosmology: theory - galaxies: formation -
galaxies: evolution - ISM - molecular processes

\section{Introduction}

The Hubble Deep Field (HDF; Williams et al.\ 1996) WFPC2 images and
recent Keck spectroscopic data (Steidel et al.~1996, hereafter S96)
indicate the presence of a population of star-forming galaxies at
redshifts as high as $\approx3.5$.  Furthermore, the HST images
indicate a narrow dispersion in morphological properties, in contrast
to the variety found in star-forming galaxies at intermediate
redshifts ($z\simeq1$). The galaxies at high redshifts have a compact
morphology, and are typically characterized by a small but resolved
spheroidal core, often surrounded by lower surface brightness
nebulosities (Giavalisco et al.\ 1996a, hereafter G96). The sizes and
scale lengths of these cores are similar to those of present day
bulges (G96). In contrast, galaxies at intermediate redshifts show a
large variety of morphologies such as very irregular objects, systems
with super-luminous star forming regions, and centrally located bright
nuclei embedded in elongated or irregular faint nebulosities
(c.f. Giavalisco et al.\ 1996b).  The physical characteristics of the
objects at high redshift are a subject of much debate. The
interpretations range from merging sub-galactic clumps (Lowenthal et
al., astro-ph/9612239) to massive disks (Prochaska \& Wolfe 1997), or
massive spheroidals (Pettini et al.~1994).

At the various redshifts, the physical structure of the Inter Stellar
Medium (ISM) allows the formation of stars.  Feedback mechanisms such
as energy input from supernovas into the ISM, heating due to ionizing
radiation, and stellar winds, play a fundamental role in regulating
the formation of stars, and therefore in the production of
metals. Madau et al.~(1996, hereafter M96) have made an important step
towards probing and understanding the metal history of the
universe. These authors have shown that the metal production rate,
which is a direct measure of the star formation rate, peaks at a
redshift between 1 and 2. A first approach in understanding the 
cosmic chemical evolution  with closed-box, inflow and outflow models
has been presented by Fall (1996; see also Pei \& Fall 1995).

The conditions necessary for the formation of a stellar population
involve the ability to cool interstellar gas and form dense molecular
clouds. Norman \& Spaans (1997, NS97) and Spaans \& Norman (1997,
SN97) have suggested that the efficient formation of stars begins with
a phase transition to a multi-phase ISM after a period of slow star
formation regulated by H$_2$ cooling.  This phase transition was
computed to occur at a metallicity $\sim 0.01Z_{\odot}$ depending on
the local star formation rate during this moderate phase.
(NS97). Furthermore, an optimal metallicity range of $\sim
0.01-0.1Z_{\odot}$ was identified, for which star formation can
proceed very efficiently (SN97). This result is independent of
geometry and is due to the absence of magnetic support for the
molecular clouds, because a low ionization fraction yields an
ambipolar diffusion time scale shorter than the free-fall time of a
molecular cloud. The opposite occurs at metallicities larger than
$\sim 0.1Z_{\odot}$, and hampers the collapse of dense clouds and
therefore the formation of stars. The optimal metallicity range can
allow the occurrence of starbursts, which can have a dramatic effect
on the evolution of dwarf galaxies, and in general leads to a period
of intense star formation in more massive systems (SN97, NS97).  For a
general population of proto-galactic disks, the peak in star formation
lies between $z=1.2$ and 2 (NS97).

The apparent evolution with redshift of galaxy morphology observed in
the deep imaging searches, and the observed redshift evolution of the
metal formation rate in the universe, must be strongly related to the
local physics of the multi-phase ISM. At each epoch, the ambient
interstellar gas of systems forming stars with a high efficiency, must
have a multi-phase structure with dense molecular clouds. As such, it
appears timely to investigate the influence of geometry and mass on
the occurrence of the transition to a multi-phase ISM, and on the star
formation history of the luminous galaxies at intermediate to high
redshifts.

\section{The Model}

The adopted model is described in detail in NS97 and SN97.  Its basic
ingredients are: 

(1) The effects of feedback on the energy budget of the interstellar
gas, due to the formation of stars and the occurrence of supernova
explosions, are included explicitly and allow the self-consistent
computation of the local star formation history.  A Salpeter (1955)
Initial Mass Function (IMF) is assumed.  A modified Schmidt law, which
includes the effects of ambipolar diffusion derived from the
ionization balance in the ambient medium, is adopted to compute the
local star formation rate. The total amount of energy injected into
the ISM by supernova blast waves is used to compute the ejection rate
of enriched interstellar gas.

(2) Prior to the phase transition, regions where the H$_2$ abundance
is large are characterized by a kinetic temperature of $\sim 1000$ K
due to vibrational line cooling (NS97).  For each mass $M$ and at each
redshift $z$, the multi-phase structure of the ISM is calculated as a
function of the ambient metallicity and of position in the
galaxy. That is, given the local gravitational pressure, the equation
of thermal balance is solved and its solutions used to determine the
density and temperature of the cold and dense, and warm and ionized,
medium.

(3) The loss of enriched interstellar material in non-spherical
systems due to blow out along the short axis of the system is taken
into account. The investigations of De Young \& Heckman (1994)
indicate that galaxies with masses $\ge 10^{11}M_{\odot}$ are robust
against complete ISM expulsion caused by the pressure wave produced by
SN explosions (blow-away). This result is independent of geometry,
unless the stellar energy depositions are in excess of $10^{59}$ erg
during a starburst phase.  In flattened systems, this process leaves
the outskirts of the ISM undisturbed, because the rarefaction wave
produced by blow-out along the symmetry axis of the galaxy will
overtake the expansion front (Figure 1 of De Young \& Heckman
1994). That is, as the pressure driving the expansion front drops, the
bubble of metal-rich hot gas stops expanding once its momentum is
dissipated.  An important consequence is that the spherical geometry
allows the metal-rich gas to travel a longer distance through the
galaxy than in the flattened case of similar mass, and leads to a
faster enrichment.

Our models do not incorporate effects such as a varying IMF, or
differential rotation.  Biasing the IMF toward more massive
stars alters the enrichment history of the ambient medium, and the
relative abundance of the $\alpha$ versus iron-peak elements.
Differential rotation might lead to a lower effective star-formation
rate in the inner regions of galaxies, due to shearing of molecular
clouds (as e.g., in our own Galaxy). 

In this Letter we study in detail the star formation history of
(proto-)galaxies with various geometries for the baryonic gas
component, and different values of the total (dark+baryonic) mass. We
consider oblate ellipsoidal systems with a baryonic mass of
$5\times10^{10}$ and $5\times10^{11}$ M$_\odot$ and a radius of $\ell
= (x^2 + y^2)^{1/2} = 8$ and 15 kpc, respectively, with axis ratios
ranging from 1 (spherical geometry) to 0.6 (morphological type
``E4'').  The baryonic density profile is assumed to be $\rho \propto
r^{-2}$, with $r^2= x^2 + y^2 + q z^2$ and $q$ the axis ratio, outside
a 1 kpc constant density core.  Constant baryonic to non-baryonic mass
ratios of $0.2$ and $0.04$ are considered. Our models are intended to
investigate the process of star formation in objects which sample the
mass distribution provided by scenarios of hierarchical structure
formation, but do not assume any specific prescription for
it. However, the masses and redshifts that we investigate are
consistent with standard CDM (see e.g., Figure 1 of NS97).

It is important to note that in massive galaxies the geometry-induced
blow-out discussed in (4) occurs at an epoch when the outskirts have
not been enriched yet. The blow-out phenomenon depends little on the
(poorly known) shape of the dark matter halo for a fixed baryonic to
non-baryonic mass ratio, given that the halos are more extended than
the baryonic matter, and is driven instead by the geometrical shape of
the baryonic mass distribution. The time scale on which blown-out gas
can be re-accreted depends on the total binding energy, i.e.~the total
(dark+baryonic) amount of matter present in the galaxy. Typical time
scales are of the order of $\sim1$ Gyr, and therefore this effect is
not likely to alter the conclusions of our work.

The spheroidal systems studied in this work are associated with
roughly spherically symmetric initial density perturbations which are
turning around and are collapsing under the influence of gravity. As
such, these systems are in quasi-equilibrium; the baryonic matter is
partially pressure supported as it settles into the dark matter halo
potential. Small deviations from sphericity will be amplified by
gravitational collapse on a free-fall time (Cole \& Lacey 1996). It is
therefore expected that a range of geometries exists for the baryonic
matter component. Any amount of non-zero angular momentum of the
spheroidals, and changes in their original shape due to contraction,
are ignored in our treatment of the local physics of star
formation. Our simulations do not imply the formation of the Hubble
sequence, nor do they require the formation of specific morphologies
at specific redshifts. Instead our simulations will show how the
physical structure of the ISM as a function of redshift, and therefore
the star formation rate in the system, depends on the total mass, and
on the morphology of the baryonic matter.

The time scales involved in the model are the following. The
(proto-)galactic systems discussed above form on a free-fall time of
$\sim 1$ Gyr due to cooling by Bremsstrahlung.  During this time, only
a small fraction of baryonic matter is converted into stars due to
H$_2$ cooling (NS97). Supernova bubbles produced in the dense central
regions expand on a time scale of $\sim 20$ Myr, and blow-out in
flattened systems. The latter time scale is shorter than the time
required to condense the metal-rich tenuous gas inside the expanding
bubbles into dense molecular clouds in the presence of star formation
($\sim 50$ Myr). Once the phase transition occurs, the bulk of the
star formation occurs on time scales much shorter than the dynamical
time scale.

\section{Results}

Figure 1 presents the volume averaged star formation rates for three
spheroidal (proto-)galaxies of morphological type E0, E2, and E4. The
solid curves represent the average over the inner 1 kpc of a galaxy
and the dashed curves correspond to the average over a shell with an
inner radius of 7 kpc and an outer radius of 10 kpc.

The initial star formation rate is driven by H$_2$ cooling, and
results in a moderate epoch characterized by a value of the local star
formation rate $\sim 10^{-2}$ M$_\odot$ yr$^{-1}$, due to the
dissociation of the H$_2$ by the stellar radiation field (NS97). The
epoch of enhanced star formation starts with the transition to a
multi-phase ISM. The three galaxies show a time delay in the star
formation peak between the inner and outer region, which is more
pronounced for flatter systems. The reason of the delay is the lack of
enrichment of the outer regions due to blow-out of metals from the
inner regions in the more oblate galaxies.  The shoulder on the rising
side of the central star formation rates (solid curves) is caused by
the rapid increase in metallicity to above the optimal upper limit of
$0.1Z_\odot$ in the very inner region on time scales shorter than the
bubble propagation time across the entire inner 1 kpc. This gives rise
to an enhancement in the magnetic support of the innermost molecular
clouds when the ambipolar diffusion time scale becomes longer than the
free-fall time.  Once the metal-rich hot phase produced during this
first burst of star formation has traversed the entire inner 1 kpc and
has been mixed with the metal-poor ambient medium, the inner star
formation rate reaches its peak. The star formation rates in both the
inner and outer regions decrease with decreasing redshift as the
available enriched molecular gas decreases because it is converted
into stars. One should note that for the galaxy sizes and density profiles
investigated here, no more than 10\% of the baryonic mass is in the cold dense
phase due to the large energy input of supernova explosions, which heat up
the interstellar gas at the star formation peak. For smaller and denser
proto-galaxies this percentage increases as the cooling time decreases.
Its relation to the early formation of massive ellipticals will be the
subject of a follow-up paper.

The massive spherical systems reach their star formation peak in the
inner and outer regions as early as $z\sim 4$, whereas more flattened
galaxies produce most of their stars between $z=2.5$ and $z=1$. This
shift is caused by the slower built-up of enriched ISM in the flattened
systems, i.e. by the loss of metals through the blow-out phenomenon.

The enriched interstellar gas is a direct measure of the total amount
of dust present in the galaxy at any time. For the E0 galaxy,
dust-to-gas ratios of 0.01 by mass or more are reached around $z\sim
4$, a value comparable to our own Galaxy. Depending on the actual
distribution of the dust, obscuration and depletion of metals into
dust grains might occur. The size of this effect is uncertain, and
will be discussed elsewhere.

In Figure 2 we show the total volume integrated star formation rate
for a sequence of spherical (E0) galaxies of different baryonic masses
(and constant baryonic to non-baryonic mass ratio of 0.2; solid
curves). The labels ``h'', ``i'' and ``l'' indicate the high ($5
\times 10^{11} M_\odot$), intermediate ($5 \times 10^{10} M_\odot$),
and low ($\le 3\times 10^9 M_\odot$) baryonic mass galaxies,
respectively. The latter curve is taken from SN97, who study the
evolution of dwarf galaxies. These curves indicate that, for a fixed
geometry, the formation of stars is delayed to later epochs for lower
mass galaxies.  Systems with a low binding energy, even though present
at high redshifts in hierarchical scenarios of structure formation, do
not sustain a multi-phase ISM at those epochs and therefore do not
form stars (SN97). Similarly, at constant baryonic mass, the
transition to a multi-phase ISM will occur at higher redshifts in
systems with larger binding energies.  This is illustrated by the
dot-dashed curve in Figure 2, which denotes the intermediate mass E0-i
galaxy discussed above, but with a baryonic to non-baryonic mass ratio
of $0.04$. The very few mass estimates available for star forming
systems at high redshift are in agreement with intermediate baryonic
masses ($\sim 4 \times 10^{10}$ M$_\odot$) embedded in $\sim 10$ times
more massive halos (Warren \& M{\o}ller 1996).  Our results are in
good agreement with the idea of ``downsizing'' with decreasing
redshift, as suggested by Cowie et al.~(1996) on the basis of their
Keck spectroscopy on the Hawaii deep fields.

The dotted curve in Figure 2 depicts the total volume integrated star
formation history of the $5 \times 10^{11} M_\odot$ proto-galactic
disks studied in NS97, i.e.\ flattened systems with axis ratios
smaller than 0.2. These systems are the limiting population of the
spheroidal (proto-)galaxies studied in our paper.  The massive disks
and E0 (proto-)galaxies bracket the redshift dependence of the star
formation history caused by geometrical effects.  It follows that the
flattened systems have had less time by $z=0$ to form stars and
consume their ISM.  Our results suggest that the high metal production
rate between $z=1$ and $z=2$ presented by M96 requires a large
fraction of the cosmological star formation to occur at these epochs
in flattened and/or low mass systems, and in disks (see also NS97).

It is evident that spheroidal (proto-)galaxies experience a period of
star formation which is much shorter when compared to the star
formation history of more flattened systems: A large fraction of the
baryonic gas is converted into stars within 1 Gyr for roughly
spherical, massive (proto-)galaxies. These are in quasi-equilibrium
and are dynamically evolving on a free-fall time. Since stars behave
as collisionless matter, the rapid conversion of interstellar gas into
stars during a quasi-equilibrium collapse phase, might naturally lead
to dissipationless collapse and therefore to velocity anisotropies as
observed in current epoch elliptical galaxies (van Albada 1982). The
final amount of rotational versus anisotropy support will depend on
the initial angular momentum of the system and the additional infall
of gaseous material. A variety of stellar systems can therefore
result.  The structure and stellar populations of the latter will be
discussed in a follow-up paper.

\section{Discussion and Conclusions}

We have shown that massive spherical galaxies can sustain a
multi-phase ISM and actively form stars as early as $z\sim 4$. Flatter
systems of similar mass reach their peak in star formation and metal
production at later epochs due to the blow-out phenomenon, which
induces loss of metals in these systems. This provides a natural
explanation for the observed galaxy morphologies at high redshifts
(G96; S96). The objects observed to form stars at redshifts $z\ge2$
can be identified with massive, (almost-)round galaxies, or with the
inner regions of massive, more flattened systems.  These inner regions
have a size of the order of the short axis of the galaxies, which
regulates the scale length for the occurrence of blow-out, and are
comparable to present day bulges. At the same redshifts, the outer
regions of flattened galaxies have not yet been significantly
enriched. These outskirts are likely to be the nebulosities around
many of the compact, spheroidal cores observed in the HDF.  The
geometrical sequence from round objects to the disks of NS97 also
provides a simple explanation for the remarkable ``migration'' of the
sites of intense star formation, from the ``bulges'' at high-$z$ to
the spiral disks at the current epoch (G96).  Mergers and possibly
dust might be partly responsible for the often observed irregular
morphologies at intermediate redshifts. Our models also suggest that
massive roundish (proto-)galaxies can sustain star formation as early
as $z\sim4.5$.  Relatively few very massive objects with a total mass
$M\ge 5\times 10^{11}$ M$_{\odot}$ will have formed by
$z\sim5$. Still, very red objects are observed at high redshifts
(c.f. Dunlop et al.\ 1996; Dickinson 1996). These might be
dust-reddened systems, or the rare relics of the earliest stages of
galaxy formation.

Since the peak in dust production coincides with the peak in the star
formation rate, massive systems contain large amounts of dust at early
epochs.  Therefore, a fraction of these systems might suffer from
obscuration. The observations by Omont et al.~(1996) of CO 5-4 and CO
7-6 emission toward a radio-quiet quasar at $z=4.69$ are particularly
interesting in this respect. They indicate the presence of a
multi-phase ISM with a rich ion-molecule chemistry which requires dust
particles for the formation of H$_2$ and subsequently more complex
species.  Investigations to assess the importance of the dust
obscuration effect will be presented in a future paper, although it
has been suggested to be small (M96).

The peak of the metal production rate of the universe, as constrained
by currently available observations (M96), occurs at $z\sim1$-$2$.
This peak coincides with the epochs of star formation of massive
proto-galactic disks of NS97 and of low-to-intermediate mass
spheroidals.  The overall shape of the cosmological star formation
rate is obtained by convolving the distribution function, which gives
the number of objects of a certain mass and a certain geometry for a
given redshift, with the intrinsic star formation histories of
galaxies of a certain mass and geometry, and by including the effects
of merger-induced star formation not treated here. Future
observations with NICMOS and ACS aboard HST promise to further improve
our knowledge of the cosmological metal formation history, and
possibly unveil substructure related to relatively faint objects such
as intermediate-mass spheroidals at their peak of star formation, or
flatter systems prior to their star formation peak. A detailed
understanding of its redshift dependence, coupled with the results of
our models for the mass and geometry of the starforming regions at
each epoch, can provide constraints to models of galaxy formation.

We thank Colin Norman, Massimo Stiavelli, and Joe Silk for
illuminating discussions. We thank the referee, Max Pettini, for his
detailed and stimulating comments. MS acknowledges with gratitude support of
NASA grant NAGW-3147 from the Long Term Space Astrophysics Research
Program.  CMC is supported by NASA through Hubble Fellowship grant
HF-1079.01-96a awarded by the Space Telescope Institute, which is
operated by the Association of Universities for Research in Astronomy,
Inc., for NASA under contract NAS 5-26555.

\newpage
\clearpage

\begin{figure}
\caption{Redshift dependence of the local star 
formation rate for galaxies of different geometries. The galaxies have
a baryonic mass of $5\times10^{11} M_\odot$ and a radius of $\ell =15$
kpc. The baryonic to non-baryonic mass ratio is fixed at $0.2$.  The
labels indicate the shape of the galaxy. E0 denotes an axis ratio of
unity, E2 of 0.8, and E4 of 0.6. Solid curves represent the inner 1
kpc of the galaxies.  The dashed curved correspond to a shell with an
inner radius of 7 kpc and an outer radius of 10 kpc.}
\end{figure}

\begin{figure}
\caption{Redshift dependence of the total volume integrated star 
formation rate for E0 galaxies of different masses. From right to left
(solid curves): $5\times 10^{11} M_\odot$ and $\ell=15$ kpc (E0-h), $5
\times 10^{10} M_\odot$ and $\ell=8$ kpc (E0-i), $\le 3\times 10^9 M_\odot$
and $\ell \le 3$ kpc (E0-l). The dotted curve represents the
proto-galactic disks of NS97 and the curve for the smallest
spheroidals is as computed by SN97. The dashed curve denotes the E0-i
model with a baryonic to non-baryonic mass ratio of $0.04$ instead of
$0.2$.}
\end{figure}

\setcounter{figure}{0}

\newpage
\begin{figure}
\centerline{\psfig{figure=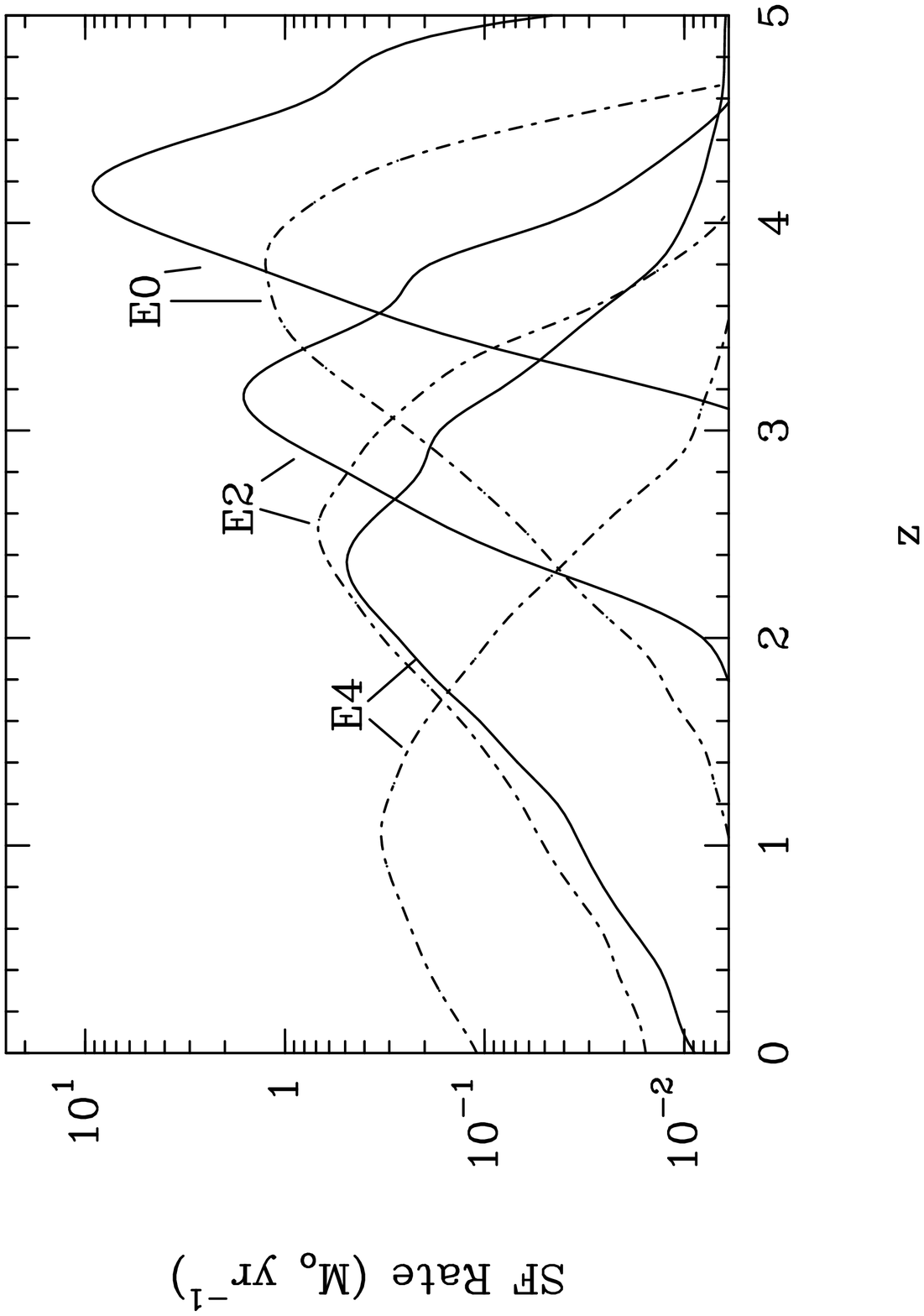,width=6in}}
\caption{\label{figure1}}
\end{figure}

\newpage
\begin{figure}
\centerline{\psfig{figure=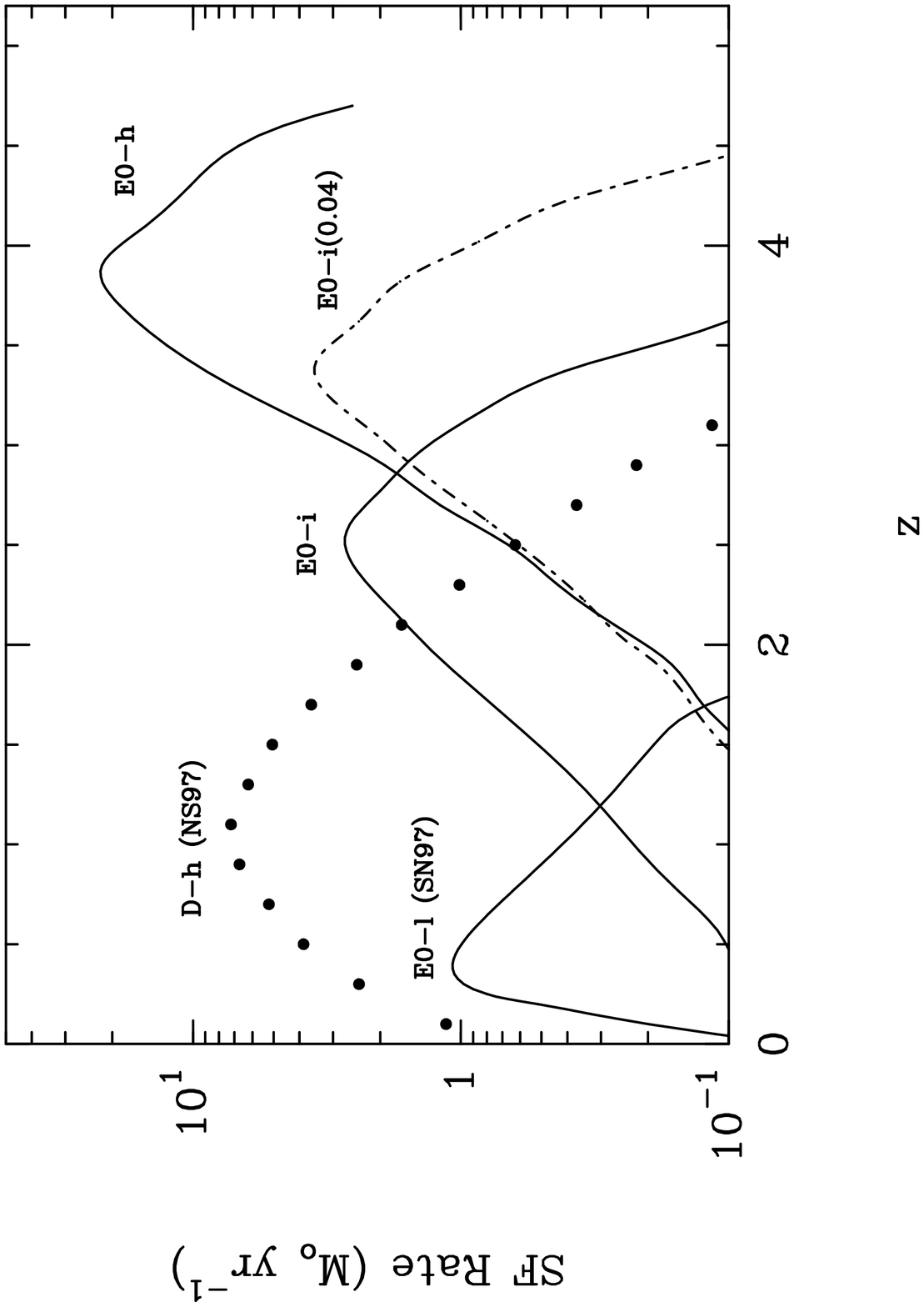,width=6in}}
\caption{\label{figure2}}
\end{figure}

\end{document}